# Localization state induced superconductivity, pseudogap and strange metal

Kuan-Ming Hung[1,*], Tung-Ho Shieh[2], and Kun-Yuan Wu[3]

[1]Department of Electronics Engineering, National Kaohsiung University of Science and Technology, Kaohsiung 807, Taiwan.

[2]Department of Intelligent Robotics Engineering, Kun-Shan University, Tainan 710, Taiwan.

[3]Division of Intellectual Property & Design Support, United Microelectronics Corporation, Tainan 710, Taiwan.

[*]Corresponding author. Email: kmhung@nkust.edu.tw

**Abstract:** The emergence of superconductivity in unconventional superconductors usually accompanies the normal-state phases of pseudogap, strange metal and Fermi liquid. It indicates these phases are strongly related to the superconducting state and should be dominated by yet unclear mechanism of superconductivity. Here, we propose a microscopic theory that, according to the coupling between localized and extended states, the pairing electrons which leap on and off a localized state result in superconducting instability. Two distinguishable pairing gaps are observed in these states, one dominates the superconducting temperature and another the temperature of a coherent pair confined at localized area. The pairing correlation between these states results in four phases. Their characteristics are manifested in universal Planckian-like resistivity and unusual carrier distribution, as observed in recent experiments.

**One-Sentence Summary:** Mechanism for cuprate superconductivity shows two pairing gaps, Planckian-like resistivity, and unusual carrier distribution.

Since the discovery of high critical-temperature ($T_c$) superconductors in 1986, the physical mechanism of superconductivity is still a mystery. The superconductivity is always accompanied by several normal-state phases, such as antiferromagnetism (AM), pseudogap (PG), strange metal (SM), Fermi fluid (FL), and charge density wave (CDW), increasing the understanding complexity and difficulty *(1-5)*. Recently, a lot of observations in linear-temperature (T) resistivity (namely strange metal) for a wide doping range are even more puzzling *(6-22)*. This character continuously extends to superconducting region under the suppression of superconductivity by applying magnetic field in Hall measurement and exhibits an unusual carrier distribution *(8-15)*. The doping evolution of linear-T resistivity nearby critical doping also reveals a quantum criticality *(11,15)*. Although, a quadratic dependence on T in resistivity is observed for the temperature greater than $T_c$ but smaller than pseudogap temperature ($T^*$) at CDW region, the suppression of CDW effect restores the linear-T resistivity of strange metal *(16)*. This indicates that the cuprate planes at SM region may concurrently exist coherent quasiparticles and Planckian dissipators *(6)*. At over-doped region, with increase of doping, the increase of non-zero residual resistivity under zero temperature and the decrease of T-dependent strength imply the existence of additional decoherent sources *(6,7)*. Furthermore, a dispersive structure of pseudogap like the largest d-wave superconducting gaps suggests that the Cooper pairs have been formed before establishing superconducting-phase order *(1)*. Apparently, the normal-state phases of unconventional superconductor are deeply relevant to the superconducting state and dominated by the mechanism of superconductivity.

In this paper, we present a microscopic theory for cuprate superconductors based on the coupling between localized and extended (itinerant) states (CLES) in $N_d$-impurity system.

In electron representation, the process of creating an itinerant Cooper pair above Fermi surface from annihilating a bound pair (Fig. 1a) or annihilating one localized electron and one itinerant electron (Fig. 1b) results in superconducting instability, named the *quantum Onnes effect* (Figs. 1c and 1d). The excitation of localized and itinerant quasiparticles from superconducting ground state establishes two pairing gaps. The pseudogap exhibits a dispersive structure as a function of chemical potential and crosses over the superconducting gap in the vicinity of optimal doping. The quantum Onnes effect enhances the pair distribution nearby the optimal doping. As well as the one-pair wavefunction leads to a universal Planckian-like resistivity, which is resulted from the itinerant electron interacted with and dragged by the localized states. Our results are consistent with recent experiments.

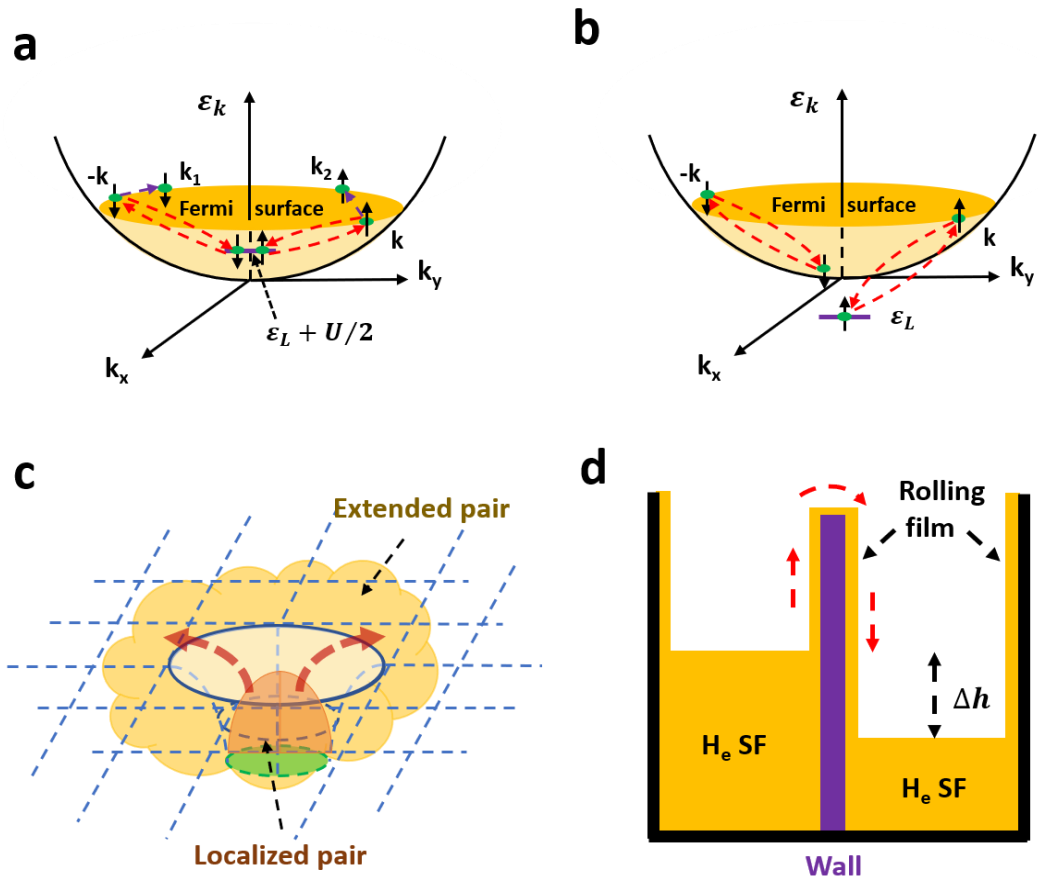

**Figure 1. Schematic of quantum Onnes effect. a.** Schematic of Cooper pair formation by the correlation of localized pair to extended pair. The purple-dashed lines indicate the extended pair breaks into normal states at $T > T_c$. **b.**

Schematic of Cooper pair formation by an antiferromagnetic electron and extended electron. **c.** Schematic of a Cooper pair creeping from a local site into extended region. **d.** Schematic of the Onnes effect in liquid-helium superfluid (SF) *(23)*.

*Model.* —— The cuprate plane in cuprate superconductors plays a pivotal role on superconductivity. In past decades, the mainstream of its theoretical description was none other than the Hubbard model. This is based on the argument that the doped carriers are populated at the copper (Cu) sites even a half of on-site Coulomb (Hubbard) repulsion U is greater than Cu d-state energy ($\varepsilon_L$) *(24)*. However, this is challenging due to the experimental observation of doped carriers populated on oxygen (O) sites *(25)* and the robust long-range superconductivity *(26)*. These observations and the Mott antiferromagnetism in cuprate superconductors suggest that Cu atom in cuprate planes is well isolated from its nearest-neighbor Cu atoms, and there is an itinerant state *(9)* coupling to the Cu localized states. Which hints that the multiple-impurity Anderson model is more suitable for characterizing the cuprate planes.

Accordingly, we propose a microscopic theory that was starting from the Anderson model with $N_d$ impurities. After canonical transformation to its Hamiltonian *(27)*, a second-order pairing approximation was made for large U. Note the spin scattering is renormalized into the zero-order kinetic energies $\varepsilon_k$ and $\varepsilon_L$ of extended and localized electrons, respectively. The spin exchange interaction is negligible for the temperature above Kondo region. As well as the hopping caused by CLES coupling is ignored under large U. For pairing ground state, the reduced Hamiltonian ($\widehat{H}_{red} = \widehat{H}_0 + \widehat{H}_2$) in analogue to Bardeen-Cooper-Schrieffer (BCS) theory *(28)* has the following form

$$\widehat{H}_0 = \sum_{k,s} \varepsilon_k \hat{c}_{ks}^\dagger \hat{c}_{ks} + \sum_\sigma \varepsilon_L \hat{d}_{is}^\dagger \hat{d}_{is} \qquad (1)$$

$$\widehat{H}_2 = \sum_{k,i} J_k \big(\hat{b}_i^\dagger \hat{a}_k + \hat{a}_k^\dagger \hat{b}_i\big) + U \sum_i \hat{b}_i^\dagger \hat{b}_i - \lambda \sum_{k,k'} \hat{a}_k^\dagger \hat{a}_{k'}.$$

Where $\hat{c}_{ks}$ and $\hat{d}_{is}$ are the annihilated fermionic operators for extended and localized electrons with the momentum k of spin s and the localized site i of spin s, respectively. $\hat{a}_k =$

$\hat{c}_{-k\downarrow}\hat{c}_{k\uparrow}$ and $\hat{b}_i = \hat{d}_{i\downarrow}\hat{d}_{i\uparrow}$ are their annihilated boson-like pairing operators. The first two terms with coupling strength J in $\hat{H}_2$ represent the processes of pairing electrons which leap on and off a localized state. The third term is the on-site Hubbard interaction. The conventional pairing interaction with strength λ is appended in the last term. The details of following derivations and numerical calculations can be found in Supplementary Materials.

*Instability.* — To study the instability of the system for the region $0 < \mu < \varepsilon_L + U/2$, we consider the process of that an extended pair above Fermi surface is created by annihilating a localized electron and an extended electron from its unpairing ground state ($|GS\rangle = \prod_{|k|<k_f,s} \hat{c}_{ks}^\dagger |AM\rangle$), where $|AM\rangle$ is the Mott antiferromagnetic ground state $|AM\rangle = \prod_{i \in g_\uparrow; j \in g_\downarrow} \hat{d}_{i\uparrow}^\dagger \hat{d}_{j\downarrow}^\dagger |0\rangle$, and $g_s$ denotes the group of s-spin sites nearest neighbor to that of $\bar{s}$-spin sites. The wavefunction can be written as

$$|\psi\rangle = \sum_{|k|<k_f,s,l\in g_{\bar{s}}} \left( \sum_{|p|>p_f} \alpha_{p,k} \hat{a}_p^\dagger + \beta_k \hat{b}_l^\dagger \right) \hat{c}_{k,s} \hat{d}_{l,\bar{s}} |GS\rangle, \quad (2)$$

that is normalized by $\sum_{|k|<k_f} \left( \sum_{|p|>p_f} |\alpha_{p,k}|^2 + |\beta_k|^2 \right) = 1/N_d$. The dynamical quantities $\alpha_{p,k}$ and $\beta_k$ are derived from applying the equations of motion ($i\hbar \frac{d}{dt} \langle GS|\hat{O}|\psi\rangle = \langle GS|[\hat{O}, \hat{H}_{red}]|\psi\rangle$) to the operators $\hat{O} = \hat{d}_{l\bar{s}}^\dagger \hat{c}_{ks}^\dagger \hat{a}_p$ and $\hat{O} = \hat{d}_{l\bar{s}}^\dagger \hat{c}_{ks}^\dagger \hat{b}_j$, respectively. $\hbar$ is the reduced Planck constant. Letting $J_p = J$ and $\varepsilon_k = \bar{\varepsilon}$ (an average kinetic energy) for simplicity, we obtain the familiar discriminant of pairing instability

$$1 = \left( J^2 \frac{N_d}{E_1} - \lambda \right) \sum_{|p|>p_f} \frac{1}{E_0 - 2\varepsilon_p}, \quad (3)$$

where $E_0 = \Omega + \bar{\varepsilon} + \varepsilon_L + U$, $E_1 = \Omega - \varepsilon_L + \bar{\varepsilon} - U$, $\bar{\varepsilon}$ has a value $\mu/2$ for a slowly varying density of states $D_0$, and μ is the chemical potential. For $J = 0$, this equation is exactly the form in BCS theory *(28)*. The sum in Eq. 3 is to be cutoff at $\varepsilon_p = \omega_c$ and converted into integral form by weighting with $D_0$. The binding energy $E_b$ is calculated by subtracting the energy $\Omega_0$, that is the solution of Eq. 3, from the energy ($\bar{\varepsilon} + \varepsilon_L$) of a localized electron plus an extended electron. Note that the discriminant of pairing instability for the region $\varepsilon_L +$ $U/2 \leq \mu$ has the same form except $E_0 = \Omega + 2\varepsilon_L + U$ and $E_1 = \Omega$. In this region, the wavefunction, that stands for the process of creating an extended pair above Fermi surface by annihilating a localized pair, has the form

$$|\psi\rangle = \sum_l \left( \sum_{|p|>p_f} \alpha_p \hat{a}_p^\dagger + \beta \hat{b}_l^\dagger \right) \hat{b}_l |GS\rangle, \tag{4}$$

normalized by $\sum_{|p|>p_f} |\alpha_p|^2 + |\beta|^2 = 1/N_d$, where $|GS\rangle = \prod_{|k|<k_f,s} \hat{c}_{ks}^\dagger \prod_i \hat{b}_i^\dagger |0\rangle$ is the unpairing ground state. The binding energy $E_b$ is calculated by subtracting the energy $\Omega_0$ from the energy $(2\varepsilon_L + U)$ of a localized pair.

The binding energy at the region $0 < \mu < \varepsilon_L + U/2$ (Fig. 2) shows that the pairing instability markedly appears in the system even at weak coupling, while, at $\mu \geq \varepsilon_L + U/2$ (Fig. S1), only appears in high coupling strength, shallow localization energy, high Hubbard potential, low chemical potential, and wide bandwidth. This apparently indicates that the localized electrons driven by Hubbard potential tend to escape from the bound state and form a Cooper pair above Fermi surface (Fig. 1c). It resembles the Onnes effect in liquid-helium superfluid (Fig. 1d) *(23)*, where the superfluid is driven by a liquid-level or temperature difference to climb over a wall barrier finding its own equilibrium, the *quantum Onnes effect*.

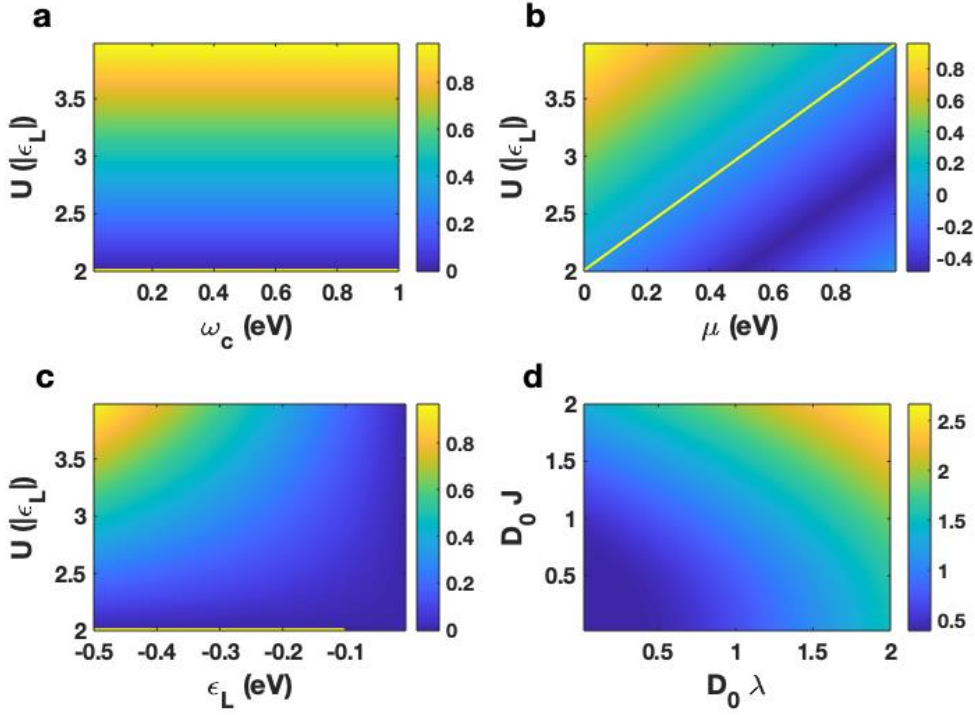

**Figure 2. Binding energy for pairing instability. (a, b, c).** Plot of $E_b$ with respect to $(U, \omega_c)$, $(U, \mu)$, and $(U, \varepsilon_L)$ measured from the energy $\varepsilon_L + \bar{\varepsilon}$ for the parameters of $D_0\lambda = 0.05$, $D_0 J = 0.09$, and $D_0/N_d = 0.804 \, eV^{-1}$, respectively. The calculated conditions are (a) $\varepsilon_L = -0.5 \, eV$ and $\mu = 0$; (b) $\varepsilon_L = -0.5 \, eV$ and $\omega_c = 1 \, eV$; and (c) $\mu = 0$ and $\omega_c = 1 \, eV$. **d.** Plot of $E_b$ measured from $\varepsilon_L + \bar{\varepsilon}$ with respect to $D_0\lambda$ and $D_0 J$ at the conditions of $\varepsilon_L = -0.5 \, eV$, $\omega_c = 1 \, eV$, $U = 2.8 \, |\varepsilon_L|$, $D_0/N_d = 0.804 \, eV^{-1}$, and $\mu = 0$. The yellow line indicates the boundary of $E_b$ from positive to negative. The units of color bar are in eV.

*Pairing ground state.* —– To characterize the pairing ground state of a system having $N$ electrons, we work with the trial wavefunction

$$|0, N\rangle = \prod_j \left(\xi_j + \varsigma_j \hat{b}_j^\dagger\right) \prod_k (u_k + v_k \hat{a}_k^\dagger) |0\rangle, \quad (5)$$

which is normalized by $v_k^2 + u_k^2 = 1$ and $\zeta_j^2 + \xi_j^2 = 1$. The number ($N_E = \sum_k v_k^2$) of extended pairs and the number ($N_L = \sum_j \zeta_j^2$) of localized pairs follow the conservation relation $N_E + N_L = N/2$. Minimizing the total energy by using the Lagrange multiplier scheme ($\delta\langle N, 0|\hat{H}_{red} - \mu\hat{N}_{op}|0, N\rangle = 0$ with the number operator $\hat{N}_{op}$ of total electrons), we have the equations $\Delta_J = J\sum_k u_k v_k$, $\chi = J\sum_j \varsigma_j \xi_j - \Delta_\lambda$ and $\Delta_\lambda = \lambda\sum_k u_k v_k$. For $N_d$ identical impurities, performing the sum with $D_0$ to these equations, yields gap equations

$$\chi = -\frac{JN_d\Delta_J}{2E_D} - \Delta_\lambda, \tag{6}$$

$$E_D = \sqrt{\Delta_J^2 + (\varepsilon_L + U/2 - \mu)^2}, \tag{7}$$

where $\Delta_\gamma = \frac{1}{2}\gamma\chi D_0 \times ln\left|\frac{-\mu+\sqrt{\chi^2+\mu^2}}{\omega_c-\mu+\sqrt{\chi^2+(\omega_c-\mu)^2}}\right|$ for $\gamma = J\ or\ \lambda$. These results can also be derived from the linearized equations of motion detailed in Supplementary Materials.

The elementary excitations of localized and extended electrons are Bogoliubov quasiparticles. The excitation energy of creating a $k\uparrow$ (or $i\uparrow$) quasiparticle is $\Gamma_{k+} = E_k$ ($\Gamma_{D+} = -U/2 + E_D$), while that of annihilating a $-k\downarrow$ ($i\downarrow$) quasiparticle is $\Gamma_{k-} = -E_k$ ($\Gamma_{D-} = -U/2 - E_D$). Notably, the minimum energy of extended quasiparticle occurs at $\varepsilon_k = \mu$. Thus, the energy gaps for localized and extended quasiparticles are $\Gamma_{D+} - \Gamma_{D-} = 2E_D$ and $\Gamma_{k+} - \Gamma_{k-} = 2\chi$, respectively. For $|\varepsilon_L + U/2 - \mu| \gg k_BT$, the critical temperatures follow the BCS-like ratio of the energy gap to its corresponding temperature $2E_G/k_BT_i = 3.51e^{2/K_0(T_i)-2/K}$, where $K = \frac{D_0J^2N_d}{2\sqrt{(\varepsilon_L+U/2-\mu)^2+\Delta_J^2}} + D_0\lambda$, $K_0(T_i) = \frac{D_0J^2N_d}{2(\varepsilon_L+U/2-\mu)}\tanh\left(\frac{\varepsilon_L+U/2-\mu}{2k_BT_i}\right) + D_0\lambda$, and $T_i$ ($E_G$) denotes $T_c$ ($\chi$) or $T^*$ ($E_D$).

The calculated T$_c$ and T$^*$ (Fig. 3) remarkably exhibit four phases as shown in Refs. *(2,12)*. The pseudogap exhibits a dispersive structure with respect to doped carrier $p = D_0\mu/N_d$ in entire region (Fig. 3a) and is underneath the superconducting gap at $p_1 \leq p \leq p_2$ as the proposed conjecture by C. Putzke et al *(8)*. The total number N$_p$ of pairs (Fig. 3b) increases at $p \geq p_1$ and saturates at $p \geq p_2$ as observed in Hall measurement *(8-10)* due to the quantum Onnes effect. The node of the pseudogap obtained by setting $\Delta_J = 0$ exhibits a quantum critical point located at optimal doping. Furthermore, the CLES coupling remarkably enhances the maximum temperature T$_c$, that occurs at the energy $\mu = \varepsilon_L + U/2$ of optimal doping, even at weak coupling (Fig. 3c).

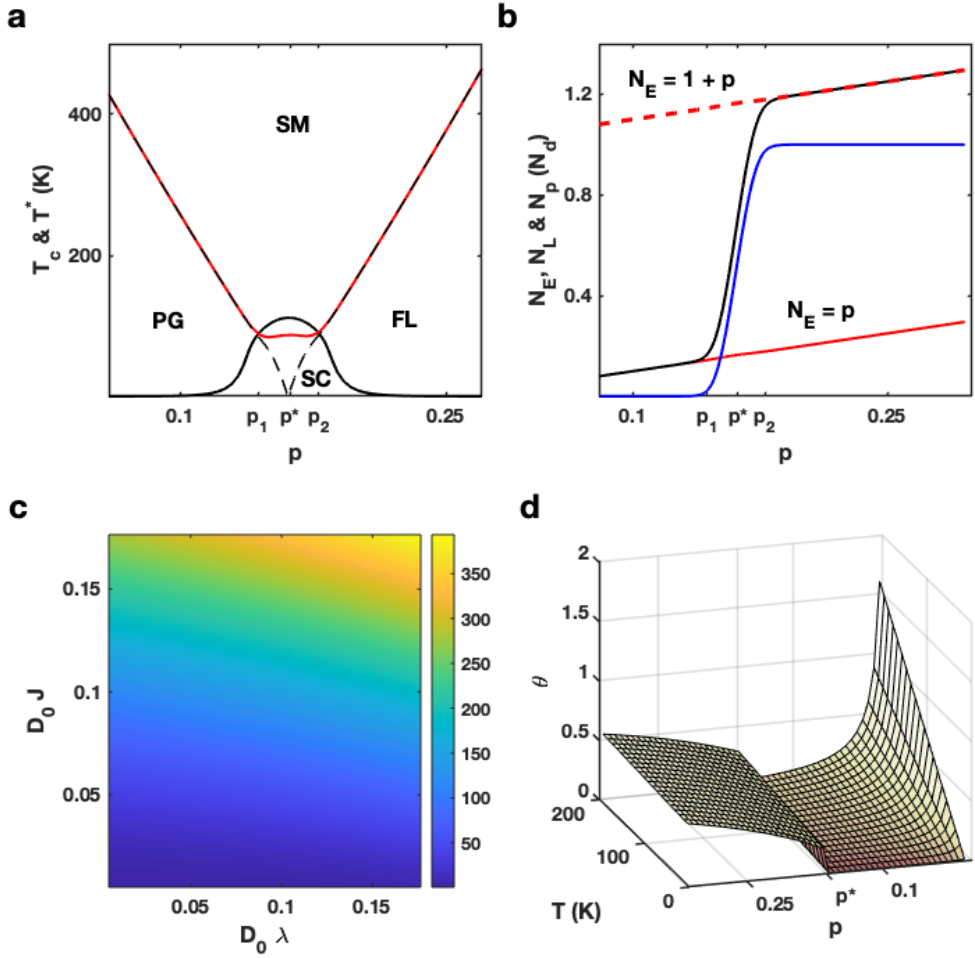

**Figure 3. Critical temperatures, carrier distributions, and resistivity. a.** Plot of calculated critical temperature $T_c$ (black line), pseudogap temperature T* (red line) and T* for $\Delta_J = 0$ (black-dashed line) with respect to doping p. There is a quantum critical point at optimal doping $p^* \cong 0.162$. The pseudogap temperature $T^*$ intersects with superconducting temperature $T_c$ at the doping $p_1 \cong 0.144$ and $p_2 \cong 0.178$. **b.** Plot of the distributions of extended ($N_E = p$, red line), localized ($N_L$, blue line) and total ($N_p$, black line) pairs. The red-dashed line shows the distribution of $N_E = 1 + p$. The calculating parameters for these two figures are $D_0\lambda = 0.05$, $D_0 J = 0.09\ eV$, $\varepsilon_L = -0.5\ eV$, $D_0/N_d = 0.804\ eV^{-1}$, $\omega_c = 1\ eV$, and $U = 2.8\ |\varepsilon_L|$. **c.** Plot of the calculated maximum $T_c$ as a function of coupling strengths at $\varepsilon_L = -0.5\ eV$, $D_0/N_d = 0.804\ eV^{-1}$, $\omega_c = 1\ eV$, and $U = 2.8\ |\varepsilon_L|$. The units of color bar are in Kelvin. **d.** Plot of the resistivity ratio as functions of temperature and doping under the parameters of $D_0\lambda = 0.05$, $D_0 J = 0.09$, $\omega_c = 1\ eV$, $\varepsilon_L = -0.5\ eV$, $D_0/N_d = 0.804\ eV^{-1}$, and $U = 2.8\ |\varepsilon_L|$.

*Resistivity.* — Since the carriers above Fermi surface are dominant in electrical transport, the electrons created from the extended pairs breaking at $T > T_c$ govern the electrical conductivity of superconductor in normal state. To characterize the phases with corresponding resistivities, the one-pair wave functions (Eqs. 2 and 4) are naturally used in the conductivity calculation by Kubo-Greenwood formula, where the transport states are weighted by the probability $|\alpha_k|^2$. Then, the conductivity has the form

$$\sigma = \sigma_0 \left[\frac{E_0/2}{k_B T} + \left(\frac{E_0/2}{k_B T} - 1\right)\frac{\mu - E_0/2}{k_B T} e^{(\mu - E_0/2)/k_B T} E_I\left(-\frac{\mu - E_0/2}{k_B T}\right)\right], \quad (8)$$

where $\sigma_0 = \frac{4e^2 \hbar \pi D_0^2 E_J^2}{3m V_g [D_0 E_J^2 + 4(\mu - E_0/2)]}$, $E_J = J N_d - \lambda E_1/J$, $E_I$ is the exponential integral function, and $V_g$ is the volume of sample. In under doping ($p < p_1$), the T-dependent resistivity ratio $\Theta = \rho/\rho_0$, the inverse of $\sigma/\sigma_0$, shows a Planckian-like (linear in T) behavior *(29)*. A minimum resistivity appears at optimal doping $p^*$. The temperature dependence of resistivity varies from strong to weak as increased doping *(17)*. The residual resistivity at zero temperature increases from zero when $p > p^*$ *(14)* and approaches one at highly doped region. The resistivity does not show a quadratic dependence on T because our theory does not consider carrier-carrier interaction and CDW effect.

*Conclusion.* — Although the CLES coupling is well known in solid state physics for a long time, the paradoxical roles of CLES coupling on impurity and localized pairing state confuse most peoples. Because the impurity is usually harmful to T$_c$ in superconductors *(30)*, and the CLES-induced pairing instability is never observed in degenerately doped semiconductor systems. To response this question, let's recall the requirements of pairing instability in the theory, including (i) well localized states, (ii) small $D_0/N_d$ ratio or high impurity density, (iii) weak hopping interaction, (iv) high Hubbard potential, (v) high coupling strength J, and (vi) wide bandwidth. However, there are two conflicting issues among them: (i) the low $D_0/N_d$ ratio and high CLES coupling imply a high wavefunction overlap between nearest-neighbor impurities, which results in the formation of impurity band, losing its locality, and reducing the Hubbard potential. The emergence of BCS superconductivity in degenerately doped semiconductors is a representative example *(31)*. (ii) The high $D_0/N_d$ ratio or diluted doping requires an unreasonable coupling strength to create pairing instability. In cuprate superconductors, the Mott antiferromagnetism in undoped cuprate planes implies a well separation between Cu atoms accompanied with high Hubbard potential due to high screening effect on Cu atom. Therefore, the pairing instability induced by CLES coupling does work in cuprate superconductors but doesn't in degenerately doped semiconductors. Based on above discussions, the theory presented in this report is a possible mechanism to the high-$T_c$ superconductors.

**Data availability**

The data that support the findings of this study are available from the corresponding author upon request.

**Acknowledgements**

This work is supported by the Department of Electronics Engineering, National Kaohsiung University of Science and Technology.

**Authors contributions**

KMH proposed the concepts and theory and wrote the article. KMH and KYW performed the numerical calculations. KMH and THS conducted the literature reviews. All authors discussed the results and commented on the manuscript.

**Competing interests**

The authors declare that they have no competing interests.